\documentclass{article}
\usepackage{arxiv}
\pdfoutput=1
\usepackage{graphicx}
\usepackage{balance}  
\usepackage{siunitx}
\usepackage{amsfonts}
\usepackage{amsmath}
\usepackage{optidef}
\usepackage{algorithm}

\usepackage{algorithmic}
\newcommand{\norm}[1]{\left\lVert#1\right\rVert}
\renewcommand{\deg}{\ensuremath{^{\circ}}}
\usepackage{subcaption}
\usepackage{bbm}

\DeclareMathOperator*{\argmax}{argmax}
\title{Rigid Graph Alignment}

\author{
  Vikram Ravindra \\
  Department of Computer Science\\
  Purdue University\\
  W Lafayette, IN 47906 \\
  \texttt{ravindrv@purdue.edu} \\
   \And
  Huda Nassar \\
  Department of Computer Science\\
  Purdue University\\
  W Lafayette, IN 47906 \\
  \texttt{hnassar@purdue.edu} \\
  \And
  David F. Gleich \\
  Department of Computer Science\\
  Purdue University\\
  W Lafayette, IN 47906 \\
  \texttt{dgleich@purdue.edu} \\
  \And 
  Ananth Grama \\
  Department of Computer Science\\
  Purdue University\\
  W Lafayette, IN 47906 \\
  \texttt{ayg@cs.purdue.edu} \\
}

\begin{document}
\maketitle

\begin{abstract}
Graph databases have been the subject of significant research and development in the database, data analytics, and applications' communities. Problems such as modularity, centrality, alignment, and clustering have been formalized and solved in various application contexts. In this paper, we focus on databases for applications in which graphs have a spatial basis, which we refer to as {\em rigid graphs}. Nodes in such graphs have preferred positions relative to their graph neighbors. Examples of such graphs include abstractions of large biomolecules (e.g., in drug databases), where edges corresponding to chemical bonds have preferred lengths, functional connectomes of the human brain (e.g., the HCP database~\cite{VanEssen13}), where edges corresponding to co-firing regions of the brain have preferred anatomical distances, and mobile device/ sensor communication logs, where edges corresponding to point-to-point communications across devices have distance constraints. When analyzing such networks it is important to consider edge lengths; e.g., when identifying conserved patterns through graph alignment, it is important for conserved edges to have correlated lengths, in addition to topological similarity. Similar considerations exist for clustering (densely connected regions of {\em short} edges) and centrality (critical edges with {\em large} weights).

 In this paper, we focus on the problem of {\em rigid graph alignment}. In contrast to a large body of work on topological graph alignment, rigid graph alignment simultaneously aligns the network, {\em as well as} the underlying structure as characterized by edge lengths. We formulate the problem and present a meta-algorithm based on expectation maximization that alternately aligns the network (using any one of the large class of existing topological aligners) and the structure (using any one of the large class of existing rigid body transformations). We demonstrate that our meta-algorithm significantly improves the quality of alignments in target applications, compared to topological or structural aligners alone. We apply rigid graph alignment to functional brain networks derived from 20 subjects drawn from the Human Connectome Project (HCP) database, and show over a two-fold increase in quality of alignment over state of the art topological aligners. We evaluate impact of various parameters associated with input datasets through a study on synthetic graphs, where we fully characterize the performance of our method. Our results are broadly applicable to other applications and abstracted networks that can be embedded in metric spaces -- e.g., through spectral embeddings.
\end{abstract}

\section{Introduction and Motivation}
Graph databases are commonly used to store data from a variety of complex systems such as financial transactions, social and communication networks, chemical reaction pathways, and biomolecular interactions. In an important subclass of such systems, the relative positions of nodes are fixed. While the network, or parts thereof, can be transformed from their preferred (relaxed) positions, their ``potential energy'' increases in the process, motivating the nodes to return to their relaxed state, i.e., the layout with the minimum energy. A natural way to think about such graphs is to visualize the edges as springs of different lengths. We refer to such graphs as {\em rigid graphs}. An intuitive example of such a graph is an abstraction of a molecule, with nodes corresponding to atoms, and edges to bonds. Bonds between specific atom pairs have well-defined lengths (e.g., a Hydrogen-Hydrogen bond has a preferred length of 74pm, a Carbon-Carbon bond, 154 pm, etc.). Bonds may be stretched or compressed, however, their potential energy increases (as determined by the Lennard Jones potential) in the process, and the molecule releases this energy in kinetic form to return to its native structure. Note that the precise interaction models (interaction potentials) are considerably more complex -- however, this high-level description is illustrative of our concept. 

To further motivate this problem and highlight its broad modeling scope, we consider common models for the human brain connectome. Brain activity can be modeled using graphs in which nodes represent regions of the brain that have specialized behavioural and/or structural properties. The coherent firing of pairs of regions is modeled by edges. In this application, the presence or absence of individual edges could vary across a population, whereas the physical positions of nodes are largely conserved in a reasonably constrained database, such as the Healthy Adult Dataset of the Human Connectome Project.

In this paper, we focus on the problem of aligning a pair of given rigid graphs. Informally stated, the goal is to find node/ edge correspondences across a pair of rigid graphs; e.g., across two given biomolecules or two brain connectomes. Traditional approaches to related problems use either structural embeddings that consider edge lengths (i.e., identifying structural transformations -- translation, rotation, dilation, that maximize overlap of graphs) or connectivity (i.e., identifying node label permutations that maximize overlap in adjacency matrices) to find correspondences between the two graphs, but not both. However, in the case of rigid graphs, both of these sources of information are rich and complementary. We demonstrate that integrating these two sources into a single framework leads to markedly better correspondences. In this paper, we formulate the problem of {\em rigid graph alignment}, and propose a solution technique for this problem in which the rigidity of nodes as well as their connectivity are used to infer significantly better alignments that either structural or topological alignment alone. 

\begin{figure}
    \centering
    \includegraphics[width=0.4\columnwidth]{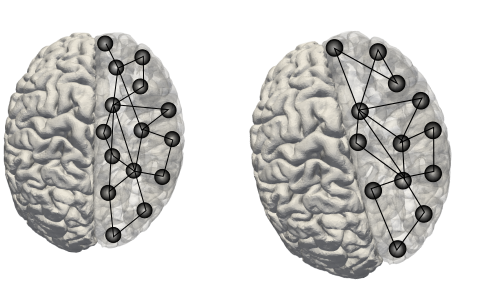}
    \caption{Illustration of two brains with functional network shown in one hemisphere. The two networks are similar in terms of node placement, and pattern of edges, but are not structurally identical -- the second brain is slightly larger and mis-oriented (in imaging), with respect to the first brain.}
    \label{fig:brain_signature}
    \vspace*{-0.2in}
\end{figure}

We motivate rigid graph alignment in two distinct application contexts. The first one uses a growing and important database of 3D brain images from the Human Connectome Project. One of the major goals of functional brain studies is to understand spontaneous firings of neurons in absence of any stimulus -- the ``resting-state'' activity. Resting state Magnetic Resonance Image (MRI) is useful in creation of a rough baseline network. A more interesting reason to study these images is that they record cognitive processes of individuals in absence of any activity. It is hypothesized that some of these cognitive processes are unique to individuals \cite{Finn15}. Thus, in principle, it must be possible to uniquely identify an individual from a population, using just their resting state functional brain image. This can be done by aligning the images structurally (through registration), topologically (through network alignment -- Figure \ref{fig:brain_signature}), or, as we demonstrate in this paper, using both structural and topological information, yielding superior results.



Our rigid graph alignment model incorporates both anatomical (structural) and functional (connectomic) information by iteratively improving on the quality of network match using a (topological) network aligner, and quality of structural match using rigid body transformations. This ensures that we: (i) do not suppress signals unique to an individual; and (ii) obtain anatomically consistent alignments. Owing to structural and functional uniqueness of individuals, we hypothesize and validate that the quality of alignment between two networks of the same individual (taken across two distinct imaging sessions) should be higher than that between two different individuals, and that ``rigid graph alignment'' is a critical methodological basis for such alignments.

We further motivate rigid graph alignment in the context of aligning large biomolecules drawn from existing databases of known drug molecules~\cite{Berman00}. This problem is of significance, since it underlies biomolecular interactions, identification of drug targets, and design of drug molecules. Specifically, binding pockets in molecules correspond to local atomic structures (active sites) -- two molecules that share active sites are likely to bind with the same ligand. This process of matching active sites can be formulated as a rigid graph alignment problem. This is an active area of research (please see Section~\ref{sec:related_research}), with commonly used methods relying on traditional graph or structural alignment, bootstrapped using techniques such as geometric hashing. These methods do not utilize constraints imposed by the bond lengths and bond angles into the graph alignment process, or the global structure of the molecules. Our novel formulation of rigid graph alignment factors the bond structure through rigid graph alignment, while simultaneously accounting for bond characteristics, using rigid body transformations.
\begin{figure}
    \centering
        \vspace{0.5cm}
    \includegraphics[width=0.3\columnwidth]{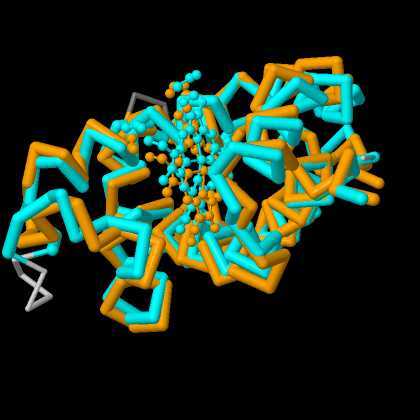}
    \caption{Structural alignment of 4HHB.A (orange) and 4HHB.B (cyan) proteins. Such alignments must account both for bond structure and bond characteristics (figure courtesy of http://www.rcsb.org \cite{Berman00})}
    \label{fig:protein_alignment}
    \vspace*{-0.2in}
\end{figure}

\textbf{Summary of Contributions} The paper makes the following contributions: (i) it mathematically formulates the rigid graph alignment problem, and puts the problem in the context of existing graph alignment and structural alignment problem formulations; (ii) it proposes a solution to the rigid graph alignment problem based on expectation maximization; (iii) it comprehensively evaluates the performance of the proposed solution on real and synthetic graphs and demonstrates significant performance improvement over state-of-the-art aligners; and (iv) it characterizes the robustness of the solution and the impact of various problem parameters through a comprehensive study on synthetic graphs.

{\em We note that our method does not compete with existing graph alignment and/or structural alignment methods. Rather, it leverages the distinct insights provided by both classes of methods to iteratively improve the quality of alignment. Our meta-algorithm can be realized with any pair of network and structural aligners to improve on the performance of either class of methods. The focus of this paper is on suitable  formulation of the problem, derivation of the meta-algorithm, and demonstration of its superiority through detailed experiments for a specific choice of structural and network aligner.}
\section{Problem Formulation}
\subsection{Problem definition}
\label{sec:formulationmethods-problemdefinition}

We define the rigid graph alignment problem by first reviewing existing graph and structure alignment formulations, and use these to motivate our new problem. 

\textit{Network alignment review.} The literature on network alignment is vast -- precluding a comprehensive review. As a representative of a broad class of techniques, we discuss the quadratic programming formulation of a network alignment objective~\cite{Bayati09}.  Let $A = G(V_A, E_A)$ and $B = G(V_B, E_B)$ be two undirected graphs that we aim to align. For simplicity, we set $|V_A| = |V_B| = n$. Let $\mathbf{L} \in \mathbb{R}^{n \times n}$ be a matrix that denotes prior knowledge on likelihood of alignment (or known similarity) between vertices $V_A$ and $V_B$. If $\mathbf{L}_{ij} > 0$, prior knowledge
implies that node $i$ in $V_A$ potentially aligns with node $j$ in $V_B$, otherwise, node $i$ in $V_A$ cannot align with node $j$ in $V_B$. 
The aim of the alignment problem is to find a matching $M$ from $V_A$ to $V_B$ using only weights of $\mathbf{L}$ and adjacency information. We define $\mathbf{X}$ to be a matrix with the same dimensions as $L$ to encode matching $\mathcal{M}$:
\[\mathbf{X}_{ij} =\begin{cases} 
      1 & \text{if\ $i \in V_A$ is matched with $j \in V_B$ under $\mathcal{M}$} \\
      0 & \text{otherwise}
   \end{cases}
\]
Under such a mapping, we say that an edge $(u,v) \in E_A$ \emph{overlaps} with an edge $(u',v') \in E_B$
if $(u,u')$ and $(v,v')$ are in $\mathcal{M}$. 

The goal in this graph alignment formulation is to find a matching $\mathcal{M}$ that maximizes a linear combination of
\emph{overlap} and matching \emph{weight}. 
Note that, if $\mathbf{X}$ is constrained by a set of matching constraints, 
\begin{equation}
    \forall i \in [|V_A|], \quad \textstyle \sum_{j} \mathbf{X}_{ij} \leq 1,\qquad \forall j \in [|V_B|], \quad \textstyle \sum_{i} \mathbf{X}_{ij} \leq 1,
\end{equation}
then
\begin{equation}
A \bullet \mathbf{X} B \mathbf{X}^T = \textstyle \sum_{ij} A_{ij} (\mathbf{X} B \mathbf{X}^T)_{ij} = \text{overlap of matching $\mathcal{M}$}
\label{eqn:edge_overlap}
\end{equation}
where we use the adjacency matrices $A$ and $B$ for the graphs and  $\bullet$ to represent a matrix inner product. 
The corresponding quadratic program is given by: 
\begin{maxi}|s| 
	  {\textbf{X}}{\alpha \mathbf{L} \bullet \mathbf{X} + \beta A \bullet \mathbf{X} B \mathbf{X}^T}{\label{eqn:netalignobj}}{}
	  \addConstraint{\textstyle \sum_{i} \mathbf{X}_{ij} \leq 1}{}{\quad \forall j=1 \ldots |V_B|}
	  \addConstraint{\textstyle \sum_{j} \mathbf{X}_{ij} \leq 1}{}{\quad \forall i=1 \ldots |V_A|, \quad \mathbf{X}_{ij} \in \{ 0, 1\} }
\end{maxi}
Here, $\alpha,\beta$ are non-negative constants that allow for tradeoff between matching weights from the prior and the number of overlapping edges. 

\textit{Review of structural alignment.} We now summarize the Orthogonal Procrustes Problem, which aims to compute structural alignments. The problem, originally solved by Sch{\"o}nemann \cite{Schoenemann66} aims to find an orthogonal transformation that reduces the distance between two matrices in the Frobenius norm. It is often used to find rigid body transformations that describe relationships between two objects. Formally, for any two matrices $Y$ and $Z$, the problem minimizes:
\begin{mini}|s|
  {\mathbf{\Omega}}{||Y - Z \mathbf{\Omega} ||_F^2}{\label{eqn:procrustes_gen}}{}
   \addConstraint{ \mathbf{\Omega}^T \mathbf{\Omega} = }{I}{}
\end{mini}
In our setting, the matrices $Y$ and $Z$ correspond to node coordinates $C_A, C_B \in \mathbb{R}^{n \times d}$. Our objective is to measure the orthogonal transformation between the sets of corresponding nodes of the two graphs. In other words, we aim to find rotation $\hat{\textbf{R}} \in \mathbb{R}^{d \times d}$ and translation $\hat{t} \in \mathbb{R}^{1 \times d}$ such that
\begin{equation}
    C_A =  C_B \hat{\textbf{R}} + \mathbbm{1} \hat{t}
\end{equation}
where, $\hat{\textbf{R}}^T \hat{\textbf{R}} = I$ and $\mathbbm{1}$ is the ones vector of length $n$. For convenience, we pad $C_A$ and $C_B$ by a vector of ones (for convenience in notation, we don't explicitly show this detail), and combine the translation vector and the rotation matrix into $\bf{\Omega}$ as
\begin{equation}
{\bf \Omega}
=
\begin{bmatrix}
\hat{\textbf{R}} &  \bf{0} \\
\hat{t} & 1
\end{bmatrix}
\end{equation}

We rephrase the problem as:
\begin{mini}|s|
  {\Omega}{||C_A - C_B \mathbf{\Omega} ||_F^2}{\label{eqn:procrustes}}{}
\end{mini}
The algorithm due to Kabsch \cite{Kabsch76} was among the first solutions aimed at solving the problem for 2D and 3D coordinates. We use a related SVD-based method, due to Sabata et al. \cite{Sabata91}, since it is shown to stable by Eggert et al. \cite{Eggert97}. 
\begin{equation}
    \mu_A = \frac{1}{n} \sum\limits_{i=1}^n C_{A_i} \hspace{10mm} \overline{C}_A = C_A - \mathbbm{1}\mu_A 
\end{equation}
\begin{equation}
    \mu_B = \frac{1}{n} \sum\limits_{i=1}^n C_{B_i} \hspace{10mm} \overline{C}_B = C_B - \mathbbm{1}\mu_B 
\end{equation}
Here, $C_{A_i}$ and $C_{B_i}$ refer to the coordinates of the $i$-th nodes of graphs $A$ and $B$. Hence, $\mu_A, \mu_B \in \mathbb{R}^{n \times 1}$. Define $\textbf{H} = \overline{C}_A^T\overline{C}_B$, then the estimated rotation matrix is $\hat{\bf{R}}$ is given by:
\begin{equation}
    \hat{\textbf{R}} = V U^T,
\end{equation}
where $V$ and $U$ are orthonormal matrices that are obtained from SVD of $\textbf{H}$.

The optimal translation $\hat{\textbf{t}}$ is defined by:
\begin{equation}
    \hat{\textbf{t}} = \mu_A - \hat{\textbf{R}}\hspace{1mm} \mu_B
\end{equation}

We note that we can estimate scaling between the two graphs using more generalized formulations of the orthogonal Procrustes problems. Instead, we assume that the coordinates are drawn from the same units (say meters). This assumption also circumvents the problem of having potentially different coordinate systems.


\textit{Rigid Graph Alignment.} We now define our problem of rigid graph alignment. 
Let $C_A \in \mathbb{R}^{n \times d}$ and $C_B \in \mathbb{R}^{n \times d}$
represent coordinates of vertices from $V_A$ and $V_B$, respectively. 
Combining the network alignment objective function and rigidity metric, as described in Equations \ref{eqn:netalignobj} and \ref{eqn:procrustes}, our objective function for {\em rigid graph alignment} can be written as:
\begin{maxi}|s|
	  {\mathbf{X, \Omega}}{\alpha \mathbf{L} \bullet \mathbf{X} + \beta A \bullet \mathbf{X} B \mathbf{X}^T - \gamma||C_A - \mathbf{X} C_B \mathbf{\Omega} ||_F^2}{\label{eqn:rigidalignobj}}{\mathbf{F} = }
	  \addConstraint{\textstyle \sum_{i} \mathbf{X}_{ij} \leq 1}{}{\quad \forall j=1 \ldots |V_B|}
	  \addConstraint{\textstyle \sum_{j} \mathbf{X}_{ij} \leq 1}{}{\quad \forall i=1 \ldots |V_A|, \quad \mathbf{X}_{ij} \in \{ 0, 1\} }
\end{maxi}
Here, we write $\mathbf{X} C_B$ to denote the conformally permuted set of coordinates of graph $B$ after alignment. 
The first term in this equation, $\mathbf{L} \bullet \mathbf{X}$, corresponds to the consistency between the prior $\mathbf{L}$ and the mapping of vertices across the two graphs $\mathbf{X}$. The second term, $A \bullet \mathbf{X} B \mathbf{X}^T$, corresponds to the alignment of the two networks, and the third term, $||C_A - \mathbf{X} C_B \mathbf{\Omega} ||_F^2$, to the structural (mis)alignment.
In {\em rigid graph alignment}, the prior $L$ is driven by spatial constraints. The weights $\alpha$, $\beta$ and $\gamma$ are parameters for the user to adjust relative importance of the prior, graph matching, and structural alignment. It can be seen that when $\alpha=1, \beta=0, \gamma=0$, the problem reduces to the maximum matching problem; when $\alpha=0, \beta=1, \gamma=0$, it is the solution to the problem of maximizing overlap; and when $\alpha=0, \beta=0, \gamma=1$, the problem reduces to rigid body transformation.  

The structural alignment error can be minimized by solving for $\mathbf{\Omega}$ using generalized Procrustes method on the ordered set of vertices. To find a good network alignment solution, we need to find the optimal $\mathbf{X}$. For ease of analysis, we view $\mathbf{X}$ as a permutation matrix. Then, the structural error term can be expressed as

\begin{align}
    &\norm{C_A - \mathbf{X} C_B \mathbf{\Omega}}_F^2 \\
    &= \text{tr}[(C_A - \mathbf{X} C_B \mathbf{\Omega})^T (C_A - \mathbf{X} C_B \mathbf{\Omega})]\nonumber \\ 
    &=C_A \bullet C_A + \mathbf{X} C_B \mathbf{\Omega} \bullet \mathbf{X} C_B \mathbf{\Omega} - 2 C_A \bullet \mathbf{X} C_B \mathbf{\Omega}\nonumber \\
    &=C_A \bullet C_A + C_B \mathbf{\Omega} \bullet C_B \mathbf{\Omega} - 2 C_A \mathbf{\Omega}^T C_B^T\bullet \mathbf{X} \label{eqn:frobenius_expansion}
\end{align}

For a given transformation matrix $\mathbf{\Omega}$, using Equation \ref{eqn:frobenius_expansion}, we can rewrite the objective function as

\begin{maxi}|s|
	  {\mathbf{X}}{\alpha \underbrace{\mathbf{L} \bullet \mathbf{X}}_{\mathclap{\text{Update Prior}}} + \beta \overbrace{A \bullet \mathbf{X} B \mathbf{X}^T}^{\mathclap{\text{Network Alignment}}} + \gamma \underbrace{C_A \mathbf{\Omega}^T C_B^T \bullet \mathbf{X}}_{\mathclap{\text{Structural Alignment}}}}{\label{eqn:rigidalignobj_restated}}{\mathbf{F} = }
	  \addConstraint{\textstyle \sum_{i} \mathbf{X}_{ij} = 1}{}{\quad \forall j=1 \ldots |V_B|}
	  \addConstraint{\textstyle \sum_{j} \mathbf{X}_{ij} = 1}{}{\quad \forall i=1 \ldots |V_A|, \quad \mathbf{X}_{ij} \in \{ 0, 1\} }
\end{maxi}

This form of the objective function suggests that the optimal $\mathbf{X}$ is one that maximizes the network alignment and structural alignment, while respecting the prior. Furthermore, it suggests that the prior $\mathbf{L}$ should be proportional to the similarity between the coordinates, i.e., $C_A \mathbf{\Omega}^T C_B^T$. However, this definition poses the potential problem of a dense prior, which can significantly increase the runtime of network alignment. To circumvent this issue, we constrain the number of non-zeros in the prior using a distance measure or number of neighbours. One such simple constraint is given by:

\begin{equation}
    \label{eqn:L_1}
    \textbf{L}_{i,j} =
    \begin{cases} 
      1 & \norm{C_{A_i} - C_{B_j}}_2^2 \leq \epsilon \\
      0 & \text{otherwise}
   \end{cases}
\end{equation}
where $\epsilon$ denotes a small neighborhood. Alternatively, we can set the $ij$-th entry of $\mathbf{L}$ to be inversely proportional to the distance between the $i$-th node of $A$ and the $j$-th node of $B$.

\begin{equation}
    \label{eqn:L_2}
    \textbf{L}_{i,j} =
    \begin{cases}
    \text{exp}(-{\norm{C_{A_i} - C_{B_j}}_2^2}) & \norm{C_{A_i} - C_{B_j}}_2^2 \leq \epsilon \\
    0 & \text{otherwise}
    \end{cases}
\end{equation}

Yet another option is to match node $i$ of graph $A$ with one of its $k$-nearest neighbours. The distance, $d^k_i$,  of $i$ and the $k$-th closest neighbor is given by:

\begin{equation}
\label{eqn:L}
    \textbf{L}_{i,j} =
    \begin{cases}
    \text{exp}(-{\norm{C_{A_i} - C_{B_j}}_2^2}) & \norm{C_{A_i} - C_{B_j}}_2^2 \leq d^k_i \\
    0 & \text{otherwise}
    \end{cases}
\end{equation}

The intuition for an rigid graph alignment technique is as follows: let $P_{ii'}$ denote the likelihood of matching node $i$ of graph $A$ with $i'$ of graph $B$. Since we use structural conformity as prior information to guide the network alignment, we say that $P_{ii'} \propto \exp(-\norm{C_{A_i} - C_{B_{i'}}}_2^2)$. We can similarly define $P_{jj'}$. Assuming that the probabilities are independent, the expected edge overlap can be written as the product:
\begin{align}
    \mathbb{E}_o & = \sum\limits_{i,j,i',j'}P_{ii'} P_{jj'} A_{ij} B_{i'j'}\\
    & = \sum\limits_{i,j,i',j'} \text{exp}(-\norm{C_{A_i} - C_{B_{i'}}}_2^2) \text{exp}(-\norm{C_{A_j} - C_{B_{j'}}}_2^2) A_{ij} B_{i'j'}
\end{align}
We can see that the expected edge overlap increases as matched nodes move closer to each other. Hence, increased structural alignment results in increased edge overlap. In an analogous manner, accurate matches lead to more accurate transformation matrices, which minimizes the residual error after structural alignment. Premised on the observation that the two objectives depend on, and reinforce each other, we propose an algorithm that optimizes the two terms, alternately as follows:

\begin{equation}
    \mathbf{X}^{(t)} = \argmax_\mathbf{\Omega} \mathbf{F}(\mathbf{X},\mathbf{\Omega}^{(t)})
\end{equation}

\begin{equation}
    \mathbf{\Omega}^{(t+1)} = \argmax_\mathbf{X} \mathbf{F}(\mathbf{X}^{(t)},\mathbf{\Omega})
\end{equation}
This formulation directly motivates our rigid graph matching algorithm in the next section.

\subsection{Rigid Graph Matching Algorithm}
\label{sec:formulation-algorithm}

Our approach to rigid graph alignment splits the problem into two tasks and iterates to convergence: (i) align the graphs restricting the prior to pairs of nodes between graphs that are in spatial proximity (using definitions for \textbf{L} mentioned earlier) -- i.e., maximizing the second term of the objective function, and (ii) align the coordinates using the current estimate of the alignment; i.e., maximizing the third term of the objective function.
This requires an initial alignment to begin, which we discuss at the end of the section. 

The vast majority of network alignment methods take as input $\alpha, \beta, L, A, B$ (or equivalent inputs). Hence, our goal in step (i) is to leverage the coordinates to estimate a matrix $L$ that constrains the set of alignments considered by the network alignment component. We devise a routine, \emph{get\_prior} that creates the matrix $\textbf{L}$ based on the distance between vertices (using the current transformation $\Omega$). Our implementation of this routine assigns every node $i$ in $A$ with $k$ of its nearest neighbours as possible matches. Each candidate node is assigned a weight inversely proportional to the distance between that node and node $i$ (Equation \ref{eqn:L}). In terms of the objective function, note that this choice of $L$ ensures that the term $||C_A - \mathbf{X} C_B \mathbf{\Omega} ||_F^2$ stays approximately the same after we use a network alignment method to optimize the permutation. 

Step (ii) involves the use of a procedure for structural alignment.  We used an SVD-based method to compute orthogonal transformation for structural alignment, as described in Section \ref{sec:formulationmethods-problemdefinition}. 

To recap, in step (i), we maximize the term $\alpha \mathbf{L} \bullet \mathbf{X} + \beta A \bullet \mathbf{X} B \mathbf{X}^T$ with the term $\gamma||C_A - \mathbf{X} C_B \mathbf{\Omega} ||_F^2$ approximately fixed, whereas in step (ii), we maximize $-\gamma||C_A - \mathbf{X} C_B \mathbf{\Omega} ||_F^2$ with the graph alignment term fixed. 

The resulting \emph{rigid graph alignment} procedure is given in Algorithm \ref{alg1} \footnote{Code will be made available with final version of the paper}.
\begin{algorithm}[tbh]
   \caption{Rigid Graph Alignment}
   \label{alg1}
\begin{algorithmic}[1]
   \STATE {\bfseries Input:} Graphs $A(V_A,E_A)$ and $B(V_B,E_B)$, Coordinates $C_A$ and $C_B$, $\alpha$, $\beta$, $\gamma$
   \STATE {\bfseries Output:} Aligned graphs $A$ and $B$
   \REPEAT
   \STATE  $\mathbf{L} = \text{get\_prior}(C_A, C_B)$
    \STATE $\mathbf{X} = \text{align}(A, B, \mathbf{L})$
    \STATE $B = \mathbf{X} B \mathbf{X}^T$
    \STATE ${\bf \Omega} = \text{transform\_coordinates}(C_A, C_B, \mathbf{X})$
    \STATE $C_B = \mathbf{X} C_B {\bf \Omega}$
   \UNTIL{converged}
\end{algorithmic}
\end{algorithm}
The process is continued until convergence is achieved, in terms of preferred metrics such as edge/node overlap or in the solution to the Orthogonal Procrustes problem. The algorithm proposed (really a meta-algorithm) works with any pair of aligners (graph and structural) that use a prior. In our experiments, we present results where we used \emph{netalignmbp} \cite{Bayati09} as the graph aligner and the SVD-based structural aligner.

In the first iteration, when we have no prior knowledge of any correspondence between nodes, we use the following approach to bootstrap algorithm \ref{alg1}: we compute average pairwise distance for all nodes and  populate the prior matrix with nodes that have similar distance profiles. We draw histograms for the distances of each node in both graphs. Then, we use correlation to measure the similarity of distance profiles of pairs of nodes across the two graphs. The prior matrix $\mathbf{L}$ is populated on the basis of this measure. We note that distance profiles are invariant to rotations and translations. Furthermore, correlation measures such as Pearson Correlation are invariant to scaling. In practice, we find that this heuristic works well both in the synthetic datasets, as well as the HCP dataset.

\section{Experimental Evaluation}

We present results from two sets of experiments -- the first set of experiments is performed on a real dataset from the Human Connectome Project. These experiments demonstrate the superiority of our method on this important dataset, while validating a number of application hypotheses. The second set of experiments is performed on synthetic datasets. These experiments are used to assess the stability of our proposed method, comparison to state of the art techniques, and demonstrating runtime characteristics.

\subsection{Analyzing the Human Functional Connectome}
We present experimental results in the context of an important application in the analyses of brain connectomes. We demonstrate that our methods are capable of significant improvement in alignment quality over state of the art graph alignment techniques that do not consider structural rigidity of the graph. 

\subsubsection{Dataset}
We construct human brain connectomes from the Young Adult Database. The Young Adult Database was created as part of the
Washington University-University of Minnesota project of the Human Connectome Project (HCP) consortium, described in
vanEssen et al. \cite{VanEssen13}. It contains both structural and behavioural data collected from over 1100 subjects,
both male and female, between the ages of 21 and 35. The images include 3T and 7T structural Magnetic Resonance Imaging
(MRI), diffusion MRI, Magnetoencephalography (MEG), Electroencephalography (EEG), functional MRI -- at resting state
and during tasks. A more detailed description of the dataset and acquisition protocol may be found in vanEssen et al. \cite{VanEssen12}
The resting state functional MRI data of each subject is collected in two sessions, which are separated by a few days.
Each session lasted 30 minutes: 15 minutes of left-to-right (LR) encoding, followed by 15 minutes of right-to-left
(RL) encoding. Each voxel is isotropic, with dimensions $2 mm \times 2 mm \times 2 mm$, and images were acquired once
every 720 ms,  as explained in
Smith et al. \cite{Smith13}. In this paper, we use only LR scans of resting state functional MRI of 20 subjects. 

\subsubsection{Preprocessing Steps}
Functional MRI data has four dimensions -- three spatial dimensions, and one temporal dimension. In essence, it
is a collection of time series data -- one per voxel of the brain. The preprocessing steps remove spatial and
temporal artifacts from the images. Head motion of subjects during acquisition is inevitable, and a potential
source of errors. We use the the motion correction tool from FMRI Standard Library (FSL), called Motion Correction
FSL's Linear Image Registration Tool (MCFLIRT) \cite{Jenkinson02}), and register each (volumetric) image in a session to the first time slice. Motion correction ensures consistent labeling of voxels between images in a session.
This is followed by skull stripping, using FSL's Brain Extraction Tool (BET) \cite{Smith02}. The images are then
resampled to voxels of size $4 mm$. This resampling is done to create networks whose dimensions are accessible to
network alignment using state of the art methods. Non-brain voxels are masked out, along with voxels of low variance.
The remaining, relevant voxels are vectorized, to create a $voxel \times time$ matrix. It has been shown that
spontaneous firings observed in resting state functional MRI is best captured in low frequency fluctuations $<$
0.1 Hz (Murphy et al. \cite{Murphy09}), which is why we use a bandpass filter with limits of 0.001 Hz to 0.08 Hz.

A $voxel \times voxel$ similarity matrix is created by computing Pearson correlation between each pair of time
series. We retain the top 5 percentile of correlation values and construct networks with voxels as nodes and
high correlation valued edges between nodes. Prior to the creation of the similarity matrix, we regress out
the global signal. Since we only use strongly positive values in subsequent steps 
we sidestep the issue of
artificially induced negative correlations.

\subsubsection{Rigid Graph Alignment Yields Higher Edge Overlap}

\begin{figure}
    \centering
    \includegraphics[width=0.5\columnwidth]{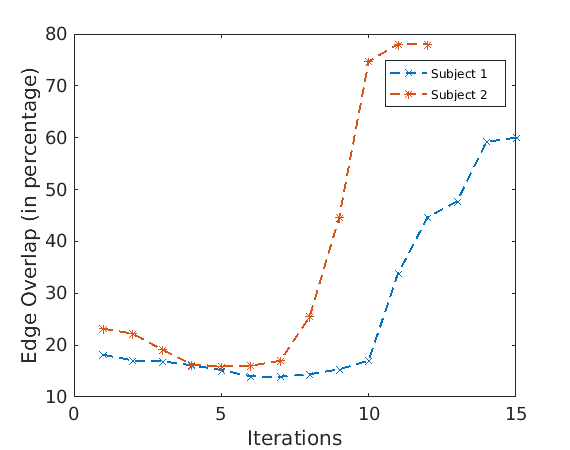}
    \caption{\textbf{Increase in edge overlap score} while aligning functional brain networks of two subjects with
    themselves (i.e., across sessions) in resting state functional MRI. These results show significant improvement
    in alignment quality over state of the art alignment techniques. Note that the alignment results after the first
    iteration correspond to that of state of the art network alignment technique. Subsequent iterations demonstrate
    improvements from our method.}
    \label{fig:intrasubject-iterations}
\end{figure}

In our first set of results, we show that Rigid Graph Alignment improves on a commonly used network alignment metric -- edge overlap. For two adjacency matrices $A$ and $B$, edge overlap is defined as $A \bullet \mathbf{X} B \mathbf{X}^T$. 
We run Rigid Graph Alignment (Algorithm \ref{alg1}) for intra-subject (same subject across two sessions) and
inter-subject (across subjects) analysis on 20 subjects, with equal weights given to the prior ($\alpha$), edge overlap ($\beta$), and structural alignment ($\gamma$), as discussed in Section \ref{sec:formulationmethods-problemdefinition} for a maximum of 20 iterations and
a convergence threshold of $0.1\%$ in edge overlap. 
We find that the edge overlap at the end of first iteration is $20.18 \pm 4.2\%$,
whereas the edge overlap after \emph{rigid network alignment} is $53.05 \pm 12.5\%$. The improvement of edge overlap is due to increasingly accurate priors in the later iterations of the algorithm.
The increase per iteration
in typical intra-subject alignment is shown in Figure \ref{fig:intrasubject-iterations}.  We observe that the
scores are largely stagnant in the first few iterations, which is attributed to the fact that the initial
prior is not as informative as priors in subsequent iterations. As stated before, to bootstrap the algorithm, we use similarity of distance profiles to populate the initial prior matrix because the ``drawings'' of graphs could look very different. In the later iterations, successive transformations result in corresponding nodes of the two graphs being placed close together.

We also characterize the statistical significance of our performance improvement. To do this,
we transform the coordinates by random transformation matrices (i.e., we randomly reorient the brains before
alignment) in 100 trials. In each case, we found that the edge overlap by rigid graph alignment was higher than edge overlap by regular graph alignment. This implies that our 
performance improvements are significant and robust.

\subsubsection{Residual Error in Structural Transformation as a Metric for Network Alignment}

The problem of identifiability of the connectome, or the so-called \emph{brain fingerprint} involves finding patterns
in brain images (either structural or functional) that are unique to an individual. In principle, two networks
belonging to the same subject should align ``better'' than two networks belonging to different subjects. In our
context, we need to devise a scoring measure for alignments that can suitably distinguish intra-subject and inter-subject alignment. 
\begin{figure}
    \centering
    \includegraphics[width=0.5\columnwidth]{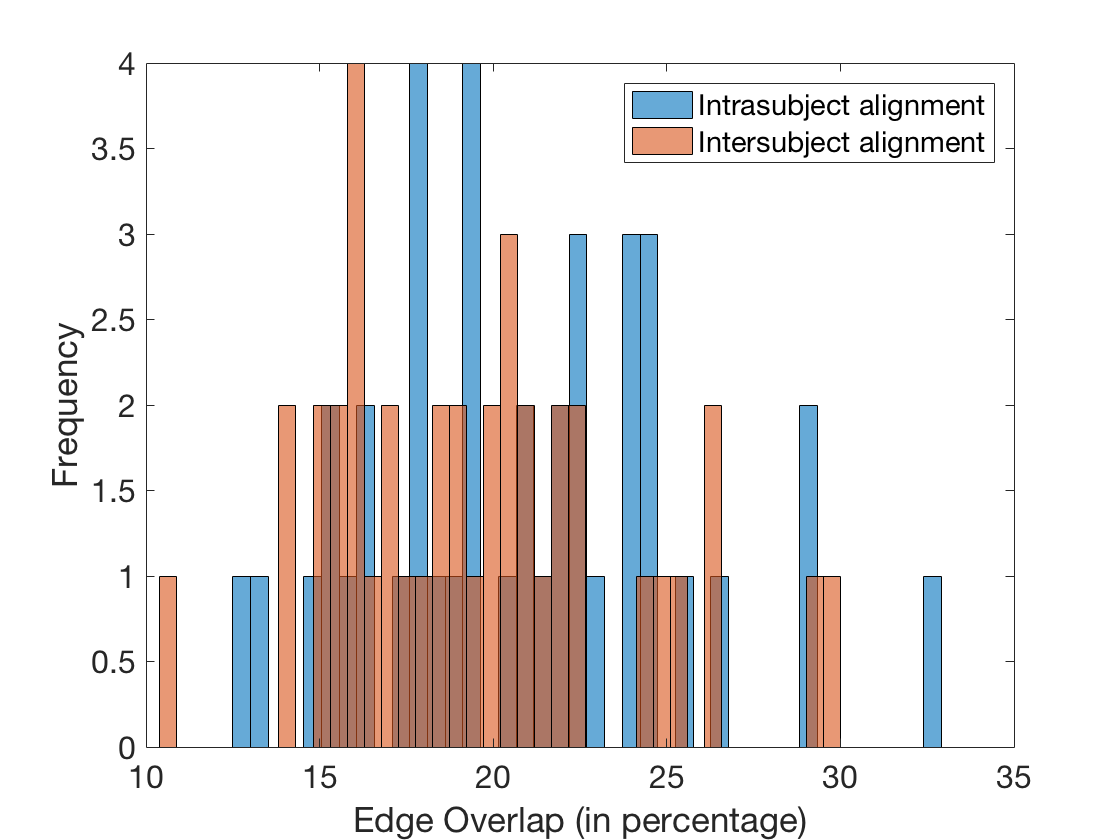}
    \caption{Histogram of edge overlap values, in both inter-subject and intra-subject alignments after first iteration. The similar overlap values show that the edge overlap is a not good metric for detecting brain fingerprints.}
    \label{fig:signature-hist-edge-overlap}
\end{figure}
\begin{figure}
    \centering
    \includegraphics[width=0.5\columnwidth]{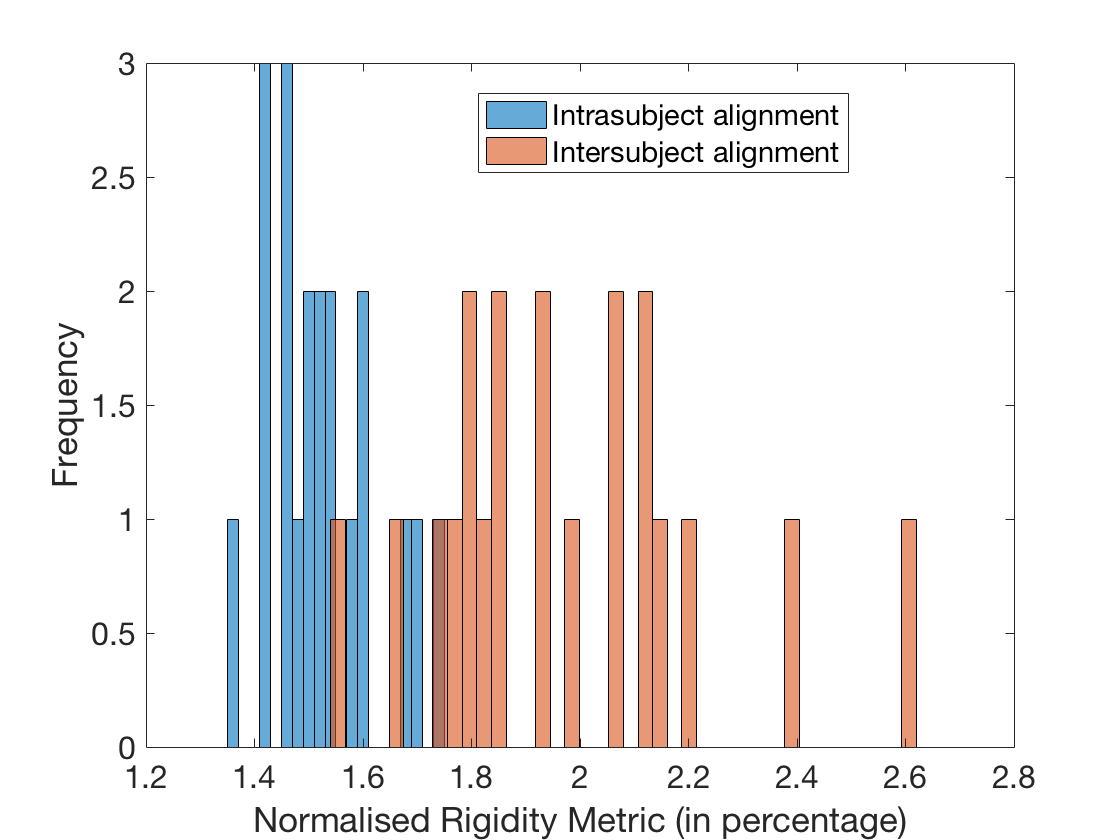}
    \caption{Histogram of rigidity values for inter-subject and intra-subject alignments. The values are significantly better separated than edge overlap, suggesting that rigidity metric is better than edge overlap in the context of brain fingerprints.}
    \label{fig:signature-rigidity-metric}
\end{figure}
\begin{figure}
    \centering
    \includegraphics[width=0.5\columnwidth]{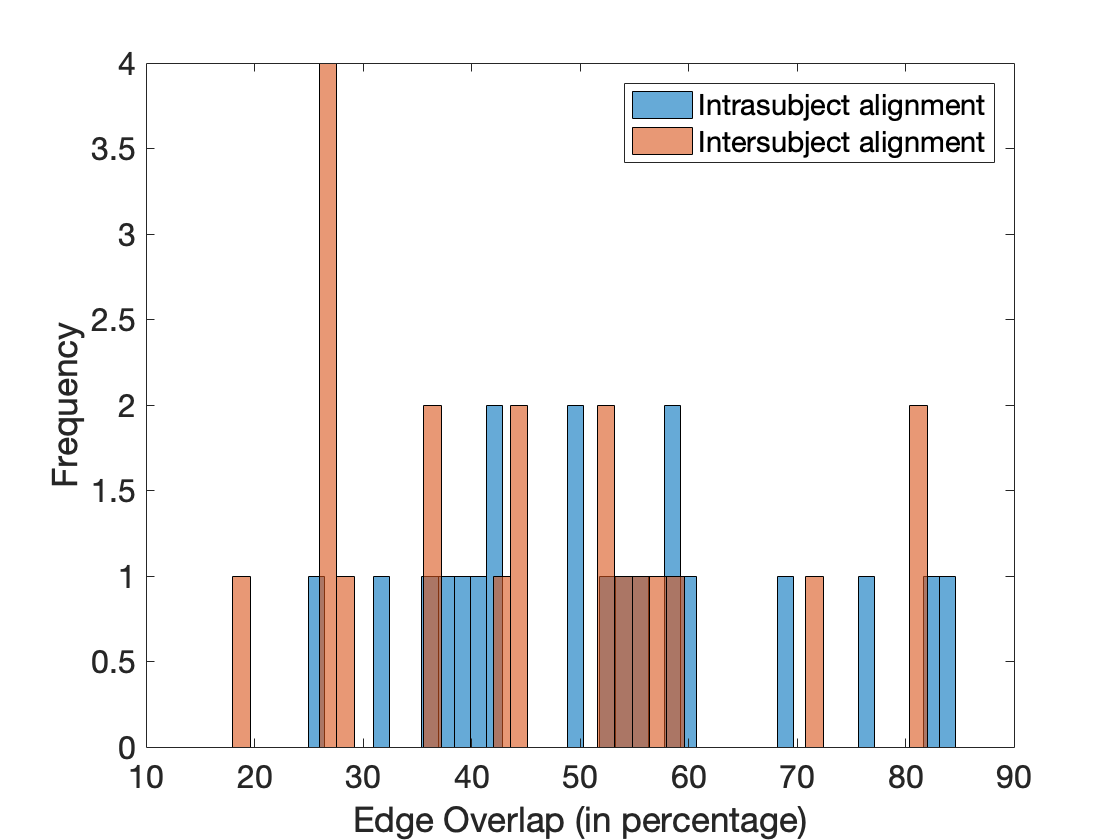}
    \caption{Histogram of edge overlap, after rigid graph alignment. The overlap has increased in both intra-subject and inter-subject alignments.}
    \label{fig:signature-histogram-edge-overlap-final}
    \vspace*{-0.2in}
\end{figure}

Figure \ref{fig:signature-hist-edge-overlap} shows the histogram of edge overlap at the end of the first iteration.
The edge overlap for intra-subject networks is $20.73 \pm 4.45\%$, whereas for inter-subject networks, it is
$19.62 \pm 4.28\%$. This suggests that edge overlap is not a strong discriminatory measure for the brain signature. In fact, the improvement
in edge overlap as a consequence of our \emph{rigid graph alignment} algorithm is observed both for
inter-subject and intra-subject alignments, as can be seen in Figure \ref{fig:signature-histogram-edge-overlap-final}. In such applications, where the position of the vertices are of importance, we show that the residual error in structural transformation, which we call ``rigidity metric'' (Equation \ref{eqn:procrustes}) is a better indicator of quality of alignment.

Figure \ref{fig:signature-rigidity-metric} shows the histogram of quality of inter-subject and intra-subject alignments. For intra-subject alignments, the \emph{rigidity metric}, normalized by the number
of vertices was found to be $1.52 \pm 0.01\%$, whereas for inter-subject alignments, it was $1.99 \pm 0.025\%$. This
demonstrates significantly higher distinguishability from the rigidity metric. 

\subsection{Experiments on Synthetic Datasets}
We test the performance of our formulation and method on synthetically generated preferential attachment and Erdos-Renyi ($G(n,p)$) graphs to assess the impact of various graph parameters. 

\textbf{Generation of Synthetic Graphs.}
Generation of synthetic rigid graphs involves two steps -- spatial location of nodes and assignment of edges between them. To locate nodes, we first create a grid that is evenly spaced in all dimensions. Each grid point has a $d$-dimensional coordinate associated with it. A grid point is assigned a node on the basis of the outcome of an independent, (biased) coin toss. 

To assign edges, we consider two strategies: Preferential Attachment \cite{Barabasi99} and Erdos Renyi ($G(n,p)$) models \cite{Gilbert59}. In Preferential Attachment or the Barab\'asi-Albert model, we start with an initial graph of $n_0$. Then, an incoming node attaches itself to a subset of existing nodes with probability proportional to their degrees, i.e., $p_i = k_i / \sum_j k_j$, where $p_i$ is the probability of attaching to node $i$, which has degree $k_i$. In the $G(n,p)$ model, an edge exists between any pair of nodes $i$ and $j$ with a probability $p$, independent of other edges.

\textbf{Perturbation Schemes}
For network alignment, we generate pairs of networks. The first graph is generated in accordance with the  aforementioned procedure. The second network is created by adding noise both with respect to the spatial location of the nodes, as well connectivity profiles of the graphs. To add spatial noise, the coordinates are rotated along each axis by $\theta\deg$. Then, each coordinate is translated by a random vector $t \in \mathbb{R}^d$. Following this, the coordinates of each node are independently perturbed by $C_n \in \mathbb{R}^{n \times d}$. Finally, edges are deleted independently with a probability $p_d$ and added with a probability $p_a$. This process yields two networks, with parametrizable differences, thereby allowing us to characterize the behaviour of both network alignment and structural alignment.


We quantify performance in terms of correctly identified edges, i.e., edge overlap (equation \ref{eqn:edge_overlap}) and node overlap (or node correctness) -- the fraction of correctly paired nodes, which corresponds to the precision for our problem. Note that this measure is applicable primarily to synthetic experiments, where the correct solution is known.

\subsubsection{Impact of perturbation to the adjacency matrix}

\label{sec:exp_pert_nodes}
\begin{figure}
    \centering
    \includegraphics[width=0.5\columnwidth]{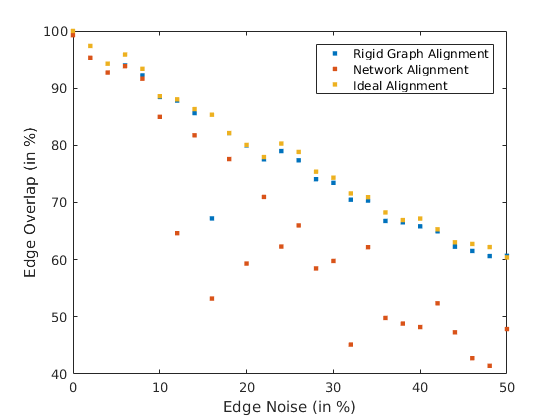}
    \caption{\textbf{Change in edge overlap with an increase in edge noise in a preferential attachment network, while node noise is fixed to zero.} It can be seen that the overlap in rigid graph alignment algorithm closely follows the true edge overlap, whereas edge overlap in regular network alignment methods is considerably lower at higher noise levels.}
    \label{fig:60_PA_1}
\end{figure}
\begin{figure}
    \centering
    \includegraphics[width=0.5\columnwidth]{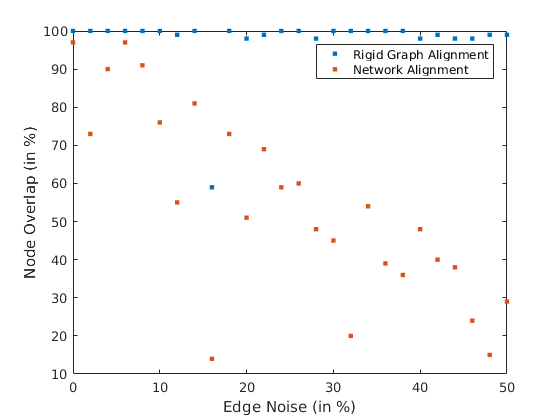}
    \caption{\textbf{Behaviour of node overlap with increase in edge noise in a preferential attachment network, while node noise is fixed to zero.} This graph shows that edge noise has no impact on the correct identification of nodes for rigid graph alignment. However, for regular network alignment, node overlap decreases linearly with increasing edge noise}
    \label{fig:60_PA_3}
    \vspace*{-0.2in}
\end{figure}

In this experiment, we add and delete edges, while holding the relative position of nodes. We allow the entire graph to be rotated along all axes. We set $p_d = p_a$ for these experiments. We repeat the experiment for $n = \{500, 1000\}$, $d = 3$, $\theta = \{1\deg, 5\deg, 30\deg, 60\deg, 90\deg, 180\deg\}$ for both preferential attachment and $G(n,p)$ graphs.

Figures \ref{fig:60_PA_1} and \ref{fig:60_PA_3} show the results of the experiment for $n=1000$, $d=3$, $\theta = 60\deg$ for preferential attachment graphs. The performance of both rigid graph alignment and a regular network alignment with respect to edge overlaps is comparable for low noise; however, rigid graph alignment is robust to higher levels of noise (Figure \ref{fig:60_PA_1}). In fact, the edge overlap in rigid graph alignment closely follows the pattern of true edge alignment. The drop in edge overlap for regular graph alignment corresponds to a drop in node alignment, as seen in Figure \ref{fig:60_PA_3}. The node overlap of rigid graph alignment is, on  average, $97.67 \pm 1.21\%$, whereas the same for regular network alignment is $65.85 \pm 3.90\%$, across different rotations. In the absence of node noise, the spatial relationships between nodes are maintained, thereby leaving edge lengths of overlapped edges unchanged. This means that a correct rigid body transformation leaves the two graphs structurally identical, which explains the high values of node overlap when rigidity of nodes is accounted for.
\begin{figure}
    \centering
    \includegraphics[width=0.5\columnwidth]{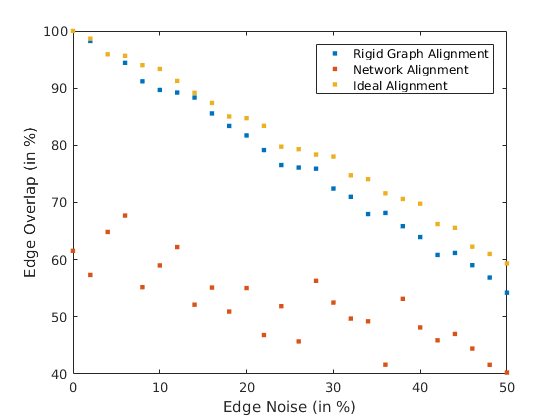}
    \caption{\textbf{Change in edge overlap with an increase in edge noise in a $G(n,p)$ graph, while node noise is fixed to zero.} The node overlap for rigid graph alignment is close to 100\% because the spatial relationship between nodes are maintained. In fact, the improvement in edge overlap is more observable even for low values of edge noise.}
    \label{fig:60_ER_1}
    \vspace*{-0.2in}
\end{figure}
\begin{figure}
    \centering
     \includegraphics[width=0.5\columnwidth]{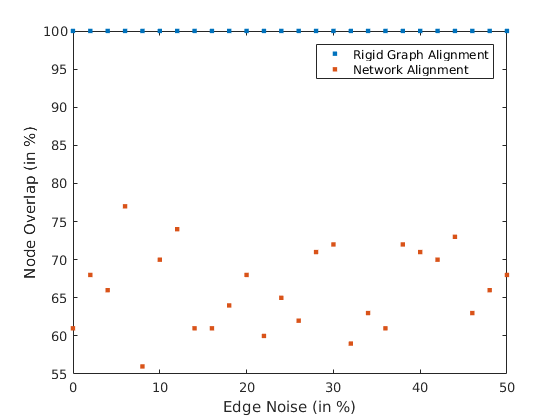}
    \caption{\textbf{Behaviour of node overlap with increase in edge noise in a $G(n,p)$ graph, while node noise is fixed to zero.} The graph shows that node overlap for rigid graph alignment is not affected by edge noise, due to the fact that rigid body transformations can transform one object perfectly into the other.}
    \label{fig:60_ER_3}
    \vspace*{-0.2in}
\end{figure}

The results of the same experiment on $G(n,p)$ graphs are presented in Figures \ref{fig:60_ER_1} and \ref{fig:60_ER_3}. Similar patterns of behavior are observed in the case of rigid graph alignment. However, the absence of patterns in edges in $G(n,p)$, coupled with the fact that regular graph aligners do not leverage structural data, causes the edge and node overlap to be substantially lower. 

\subsubsection{Impact of perturbation by node movement}

\label{sec:exp_pert_edges}
\begin{figure}
    \centering
    \includegraphics[width=0.5\columnwidth]{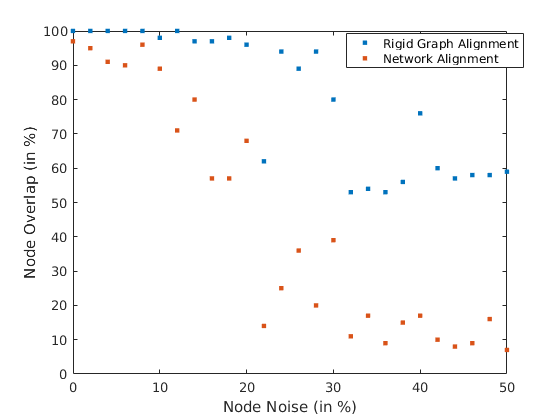}
    \caption{\textbf{Relationship between node overlap and node noise with zero edge noise in a preferential attachment network.} Large perturbation to positions of nodes lead to inaccurate priors, which explains the decrease in node overlap. However, rigid body alignment increases similarity between the networks, which explains the difference in node overlaps between the two methods.}
    \label{fig:60_PA_4}
\end{figure}
\begin{figure}
    \centering
    \includegraphics[width=0.5\columnwidth]{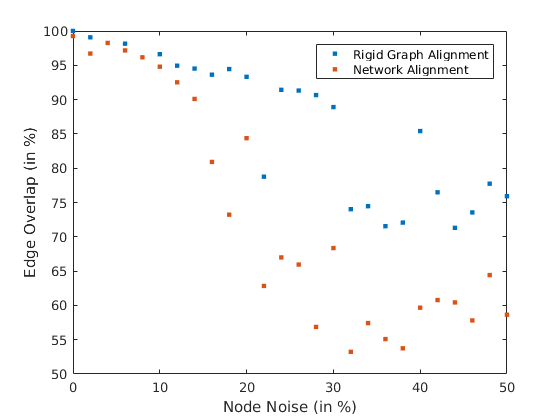}
    \caption{\textbf{Relationship between edge overlap and node noise with zero edge noise in a preferential attachment network.} The drop in node overlaps corresponds to the drop in edge overlap.}
    \label{fig:60_PA_2}
\end{figure}

In this experiment, we perturb the physical position of nodes, while fixing the edge noise to zero, i.e., $p_a = p_d = 0$. The nodes are perturbed along each coordinate by independently sampled, scaled random matrix $C_n$. The scaling is done so as to represent the perturbation as a proportion of the physical size of the network. The experiment was done for $n = \{500, 1000\}$, $d = 3$, $\theta = \{1\deg, 5\deg, 30\deg, 60\deg, 90\deg, 180\deg\}$ for both preferential attachment and $G(n,p)$ graphs.

The results for $n=1000$, $d=3$, $\theta=60\deg$ in preferential attachment graphs is shown in Figures \ref{fig:60_PA_4} and \ref{fig:60_PA_2}. In rigid graph alignment, the node overlap is $80.35 \pm 1.21\%$, whereas the same for regular network alignment is $35.85 \pm 3.90\%$. The decrease in node overlap for high node noise is due to the fact that large perturbations to individual nodes decrease the efficacy of the correction made by rigid body transformations. However, the transformations still make the graphs structurally similar, which explains the significant difference in node overlap between the two methods.
\begin{figure}
    \centering
    \includegraphics[width=0.5\columnwidth]{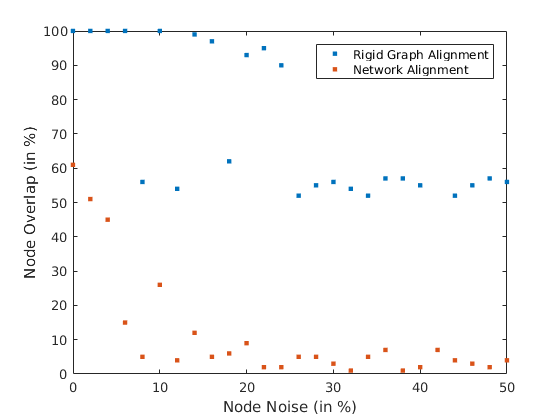}
    \caption{\textbf{Node overlap for different perturbations in node positions in a $G(n,p)$ network.} The rigid graph alignment algorithm recovers most nodes for low noise, whereas both algorithms perform poorly with higher degrees of node noise.}
    \label{fig:60_ER_4}
\end{figure}
\begin{figure}
    \centering
    \includegraphics[width=0.5\columnwidth]{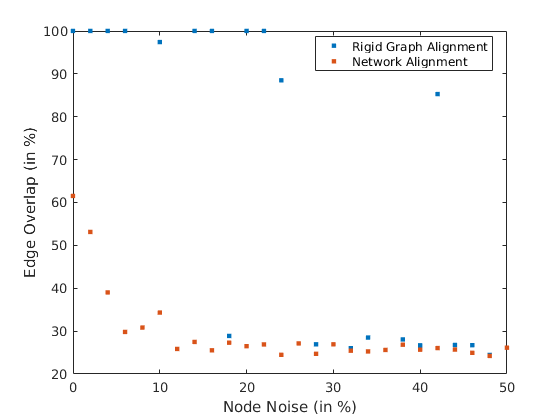}
    \caption{\textbf{Edge overlap for different perturbations to node positions in a $G(n,p)$ network.} The high node overlaps correspond to high edge overlap. However, as soon as the node overlap drops below 80\%, edge overlap has a significant drop}
    \label{fig:60_ER_2}
    \vspace*{-0.2in}
\end{figure}

Figures \ref{fig:60_ER_4} and \ref{fig:60_ER_2} show results for the same experiment on $G(n,p)$ graphs. These figures show that an absence of patterns in edges of a $G(n,p)$ graph, coupled with poor priors due to perturbations in node positions make both algorithms more sensitive to noise.

\subsubsection{Impact of perturbation of both nodes and edges}
\label{sec:exp_pert_nodes_edges}
\begin{figure*}
\begin{subfigure}{.5\textwidth}
  \centering
  \includegraphics[width=.6\linewidth]{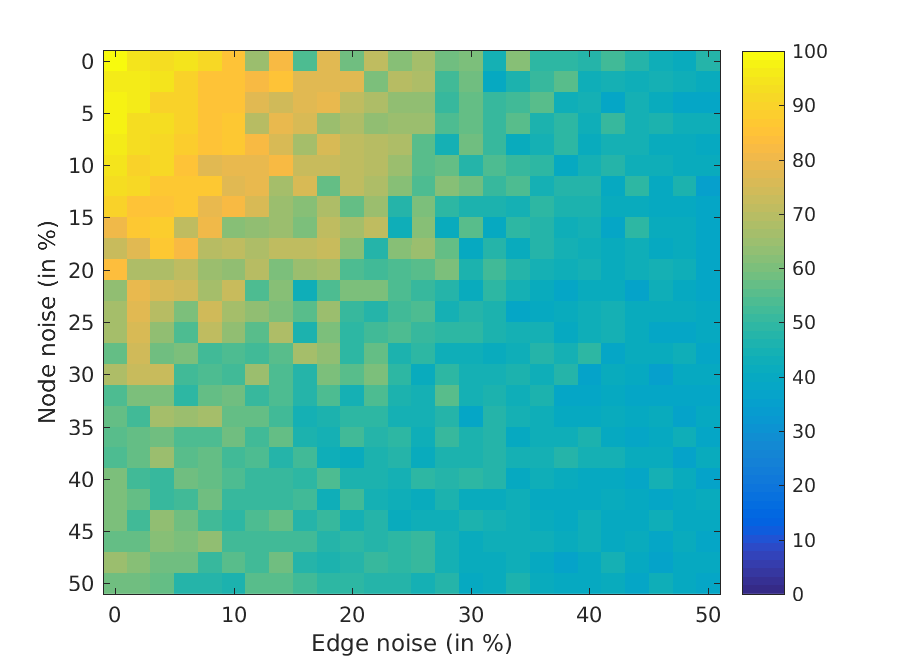}
  \caption{}
  \label{fig:PA_EO_regular_alignment}
\end{subfigure}%
\begin{subfigure}{.5\textwidth}
  \centering
  \includegraphics[width=.6\linewidth]{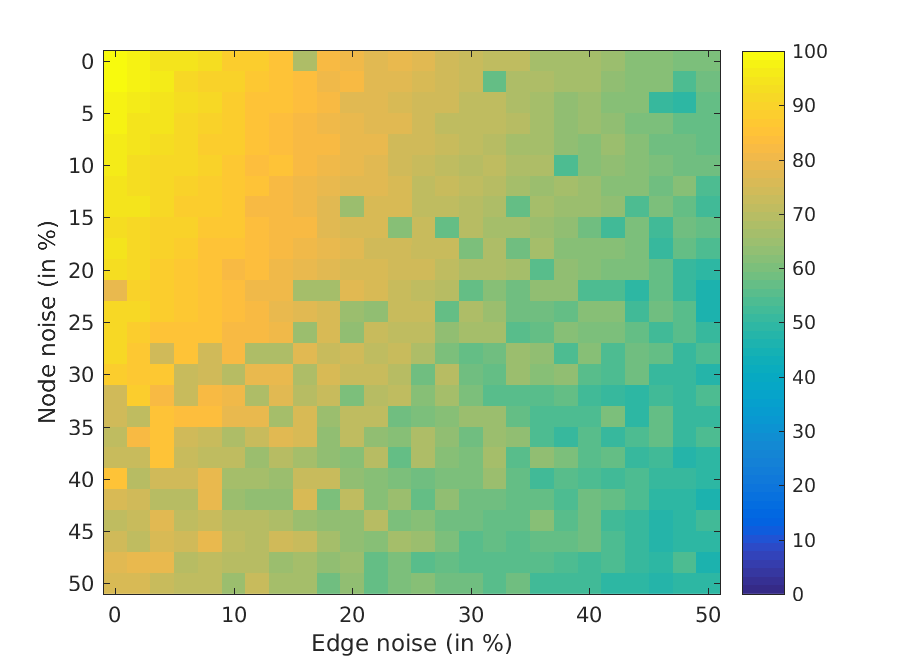}
  \caption{}
  \label{fig:PA_EO_rigid_alignment}
\end{subfigure}
\caption{Heatmap showing edge overlaps for various degrees of edge and node perturbations in a preferential attachment network with (a) regular network alignment and (b) rigid graph alignment. In regimes with low noise, the edge overlap of a regular network aligner matches the rigid graph aligner. However, at moderate noise levels, the number of edges recovered falls sharply. However, in rigid graph alignment, we are able to recover most edges. Even with higher edge noise, the edge overlap closely matches the true edge overlap.}
\end{figure*}

The effect on edge overlaps in preferential attachment networks of perturbing both nodes and edges is shown in Figures  \ref{fig:PA_EO_rigid_alignment} and \ref{fig:PA_EO_regular_alignment}. In rigid graph alignment (Figure \ref{fig:PA_EO_rigid_alignment}), edge overlaps are more stable with moderate amounts of both edge and node noise. Edge overlap values track true edge overlap. Edge overlap in the case of regular network alignment, however shows a sharp decline for small amounts of noise. Note that the scales on the two figures are made identical to facilitate comparison.

\begin{figure*}
\begin{subfigure}{.5\textwidth}
  \centering
  \includegraphics[width=.6\linewidth]{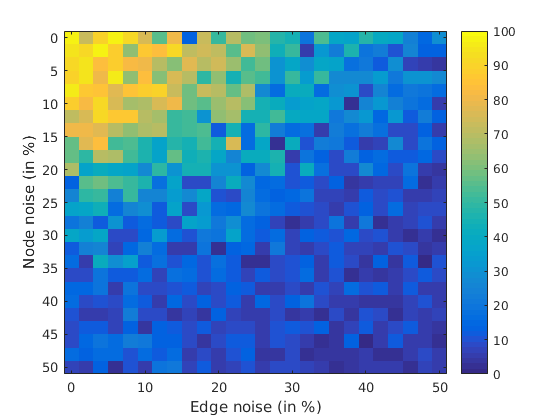}
  \caption{}
  \label{fig:PA_NO_regular_alignment}
\end{subfigure}%
\begin{subfigure}{.5\textwidth}
  \centering
  \includegraphics[width=.6\linewidth]{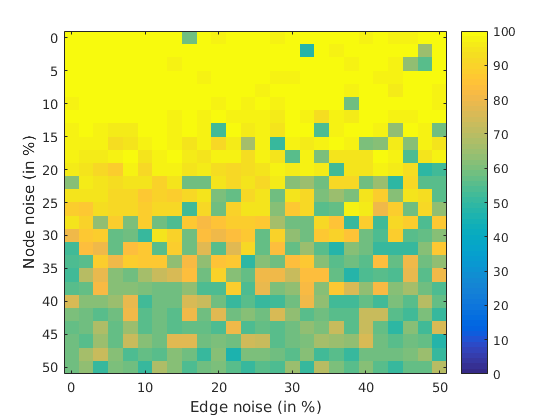}
  \caption{}
  \label{fig:PA_NO_rigid_alignment}
\end{subfigure}
\caption{Heatmap showing node overlaps for various degrees of edge and node perturbations in preferential attachment network with (a) regular network alignment and (b) rigid graph alignment. In regular network alignment, node overlap is affected by increasing values of either or both noise models. However, rigid graph alignment is much more robust}
\end{figure*}

The effect on node overlaps in the same setting is shown in Figures \ref{fig:PA_NO_rigid_alignment} and \ref{fig:PA_NO_regular_alignment}. Node alignment in rigid graph alignment is determined primarily by the amount of noise in the location of the nodes. However, regular alignment is far more sensitive to either noise.

\subsubsection{Choice of graph aligners}

\begin{figure*}
\begin{subfigure}{.5\textwidth}
  \centering
  \includegraphics[width=.6\linewidth]{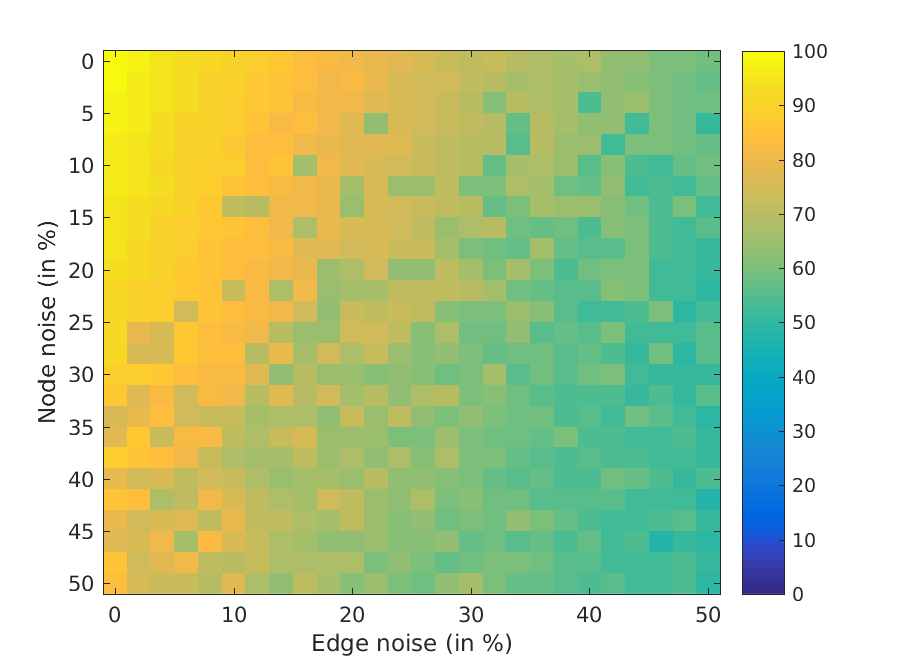}
  \caption{}
\end{subfigure}
\begin{subfigure}{.5\textwidth}
  \centering
  \includegraphics[width=.6\linewidth]{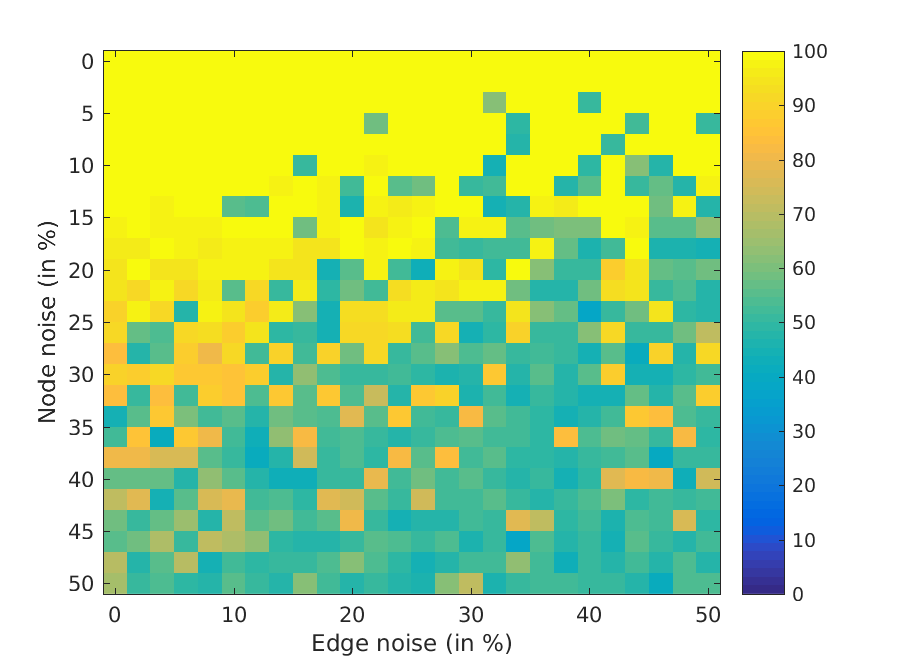}
  \caption{}
\end{subfigure}
\caption{Heatmap showing (a) edge overlap and (b) node overlap while using Klau et al. \cite{Klau09} for network alignment. The edge overlap obtained here is comparable to the belief propagation algorithm. For higher values of edge noise, the node overlap is lesser than what was observed for the belief propagation algorithm.}
\label{fig:netalignmr_PA}
\end{figure*}

\begin{figure*}
\begin{subfigure}{.5\textwidth}
  \centering
  \includegraphics[width=.6\linewidth]{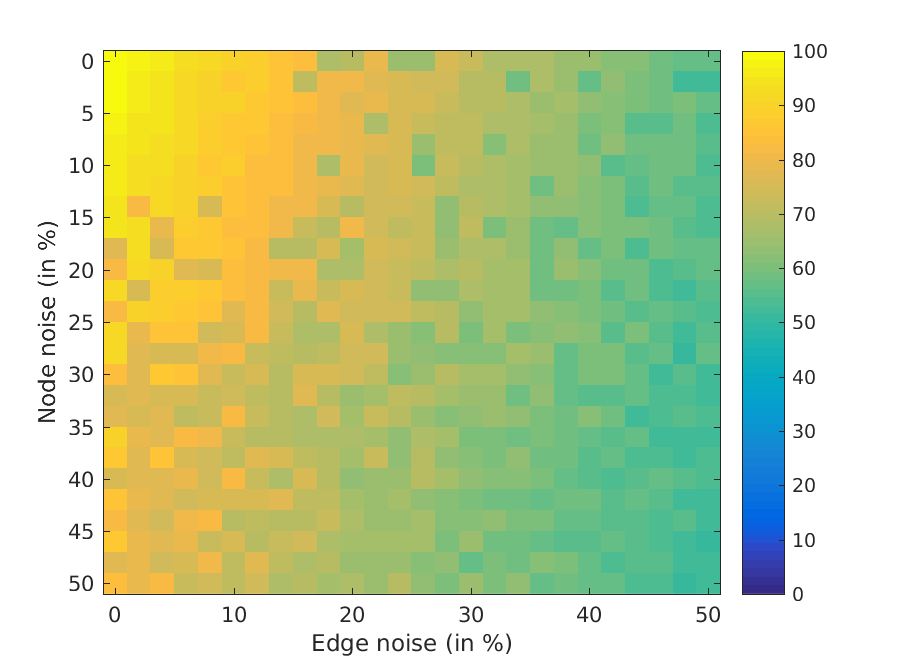}
  \caption{}
\end{subfigure}
\begin{subfigure}{.5\textwidth}
  \centering
  \includegraphics[width=.6\linewidth]{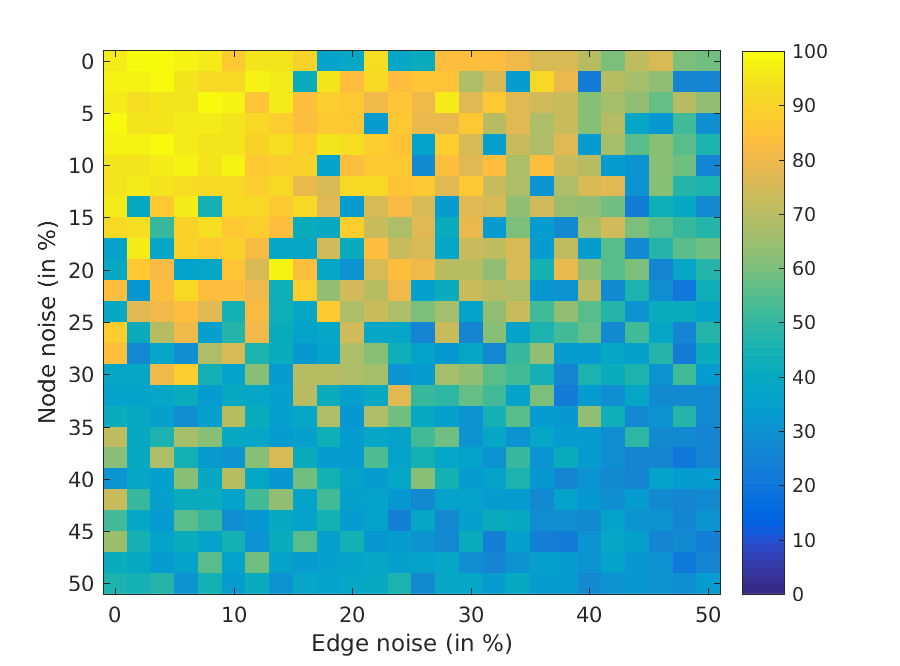}
  \caption{}
\end{subfigure}
\caption{Heatmap showing (a) edge overlap and (b) node overlap while using Isorank as aligner for rigid graph alignment. The edge overlap obtained here is comparable to the belief propagation algorithm, however the node overlap suffers significantly with edge noise.}
\label{fig:isorank_PA}
\end{figure*}

In this section, we compare the performance of the previously used belief propagation algorithm (Figures \ref{fig:PA_EO_rigid_alignment} and \ref{fig:PA_NO_rigid_alignment}) with the method due to Klau et al. \cite{Klau09} in Figure \ref{fig:netalignmr_PA} and Isorank \cite{Singh07b} in Figure \ref{fig:isorank_PA} on preferential attachment networks. As before, we vary edge and node noise and record the edge and node overlap in each case. 

We see that the algorithm due to Klau et al. \cite{Klau09} is comparable with belief propagation in terms of edge overlap, however it does slightly worse in terms of node overlap when the edge noise is high ($> 20\%$). Similarly, Isorank performs quite well in terms of edge overlap, however the node overlap degrades sharply with edge noise. Based on these observations, we conclude that our meta-algorithm can indeed be used with a wide variety readily available network aligners. 

\subsection{Runtime Considerations}

Our rigid graph alignment method is an iterative process, where a network aligner is called in every iteration. As such, it can be used in conjunction with any aligner that accepts a prior. Hence, the runtime depends primarily on the base alignment technique, and is linear in the number of EM steps. Our choice of the network alignment algorithm -- \emph{netalignmbp} \cite{Bayati09} takes $\mathcal{O}(nnz(A \otimes B) + |E_L| + matching)$, where $E_L$ is the number of edges in the prior $L$. Here, $\mathcal{O}(matching)$ is the complexity of bipartite matching (Khan et al.\cite{Khan12}), which depends on the matching algorithm used. In typical experiments, as we have shown, a small number of EM steps yield significant improvement in solution quality.

\section{Related Literature}
\label{sec:related_research}

\textbf{Network alignment} has been an active area of research over several decades \cite{Emmert16}.
Early formulations of network alignment, in the form of exact graph matching (subgraph isomorphisms)
in small chemical networks were analyzed by Sussenguth \cite{Sussenguth}. The notion of distance between
graphs (or their similarity) was first quantified by Zelinka \cite{Zelinka75, Zelinka92} on the basis
of isomorphisms, assuming that the graphs have the same number of nodes. An analytical solution to approximate weighted graph matching problem was explored by Umeyama \cite{Umeyama88}. This method used eigen-decompositions of weighted adjacency matrices to find qnear optimum matches using the Hungarian Method.

For larger graphs, inexact techniques are used for matching graphs, due to the high computational cost of
isomorphism-based methods. These formulations largely fall into two classes -- local
aligners and global aligners. Local aligners (e.g., AlignNemo \cite{Ciriello12}) aim to find
localized regions in graphs that align well. A localized network alignment approach was used by L{\"a}ssig et al. \cite{Lassig04} to align gene regulation networks in \emph{E. coli}. The optimization criteria in this class of
methods rewards local matches and does not penalize mismatches over entire graphs.
Global aligners, on the other hand, use optimization
functions that reward matches and penalize mismatches over the entire graph. 
In \cite{Singh07a, Singh07b}, Singh et al. describe IsoRank, a pairwise global network alignment method
for protein interaction networks. IsoRank is based on the intuition that a vertex in the first network should
be matched with a vertex in the second network if and only if the neighbors of the two vertices
are also well matched. A number of subsequent results have extended IsoRank, including approximations
by Kollias et al. \cite{Kollias12} and multiple graph alignment in IsoRankN by Liao et al. \cite{Liao09}.
While IsoRank is formulated as a Quadratic Program (QP), there are linear relaxations used
in network alignment by Klau et al. \cite{Klau09}, which has quadratic memory requirements. In addition to the aforementioned IsoRankN, other examples for multiple network alignment include \cite{Heimann18, Vijayan18, Nassar18}.

Many memory-efficient heuristic methods such as GRAAL \cite{Kuchaiev10} and GHOST \cite{Patro12} have been proposed, which require cubic runtime. A principled approach, based on low rank approximation of the similarity matrix used in Isorank was described by Mohammadi et al. \cite{Mohammadi17}, which does not require quadratic memory. Another method that relies on eigenvectors, called EigenAlign was introduced by Feizi et al. \cite{Feizi16}. EigenAlign estimates the similarity matrix using the dominant eigenvector of a matrix related to the product-graph between the two networks. 

Many alignment techniques incorporate prior knowledge of potential matches
to restrict the search space of potential matches to a smaller subset of nodes, and to
guide the search process. Examples of such algorithms include the previously described Isorank \cite{Singh07a}, where sequence similarity of proteins is often used as a prior for matching networks of interacting proteins. A message passing algorithm \emph{netalignmbp},
proposed by Bayati et al. \cite{Bayati09} is based on making greedy decisions constrained by a given prior, assuming that there are no cycles in the graph.

In this paper, we focus on global
graph alignment techniques that incorporate prior knowledge to yield a one-to-one mapping of nodes. 
We use \emph{netalignmbp} with an informative prior as our network alignment substrate. However,
our algorithm can be used in conjunction with any network alignment technique that allows the
use of a prior. 

\textbf{Network alignment in Graph databases} The problem of graph alignment is also commonly applied to databases. Melnik et al. \cite{Melnik02} introduce a meta algorithm to match different data structures, or \emph{models}. These models are modeled as graphs, and network alignment algorithms are used to find matching nodes. In this application, network alignment is used to manipulate and maintain schemas and match results. However, human intervention is required to validate the matches. In \cite{Buneman16}, the problem of alignment between versions of evolving RDF databases was explored. In particular, the paper proposes a method to align nodes of two consecutive versions of an RDF database using a bisimulation-based approach. The approach is flexible to accommodate nodes without labels (called ``blank'' nodes). In another network alignment algorithm for databases, Zhang et al. \cite{Zhang16} define constraints for \emph{consistency} and align nodes accordingly. 

In \cite{Heimann18a}, the authors introduce a greedy approach to identify node matchings by aligning their latent feature representations. In the first step, graph attributes such as degree and neighbors are used to establish the identity of nodes. Following this, a similarity matrix is computed for pairs of nodes across the two graphs on the basis of the previously established attributes. Then, nodes are matched in a greedy fashion. The problem of alignment of social networks with location data was explored by Riederer et al. \cite{Riederer16}. While these methods use node and/or edge attributes to construct similarity measures between nodes, they are not well suited for the applications that motivate our approach. We explicitly leverage the fact that positions of nodes are rich sources of information, and that their geometric transformations improve the quality of alignment.

\textbf{Structural Alignment} One of the commonly used approaches to \emph{structural alignment} uses the \emph{iterative closest
point} (ICP) method of Besl et al. \cite{Besl92}, and its variants \cite{Bouaziz13, Chetverikov02}.
The generic steps of ICP described in \cite{Rusinkiewicz01} for matching two structures involve selection of
source points from both bodies. ICP then matches the sets of points and computes weights corresponding
to the matches appropriately. Points that do not match are rejected and the transformation that minimizes
error is computed. This process is repeated to convergence. Our algorithm can be thought of as loosely following this framework, except that we use network alignment as the process to find suitable matching. The transformation itself, as stated in Equation \ref{eqn:procrustes_gen}, is called the Orthogonal Procrustes Problem \cite{Schoenemann66}. Kabsch's algorithm \cite{Kabsch76} provides an efficient solution to this problem in three dimensions. In fact, the graph matching problem as stated by \cite{Umeyama88} can be thought of as a two-sided Procrustes problem (Conroy et al. \cite{Conroy11}), who approximate graph matching when vertices of the two graphs can be parcellated into groups.

\section{Concluding Remarks}

This paper formulates a novel and important problem of aligning a class of graphs called rigid graphs. The problem integrates topological and structural alignments into a single framework, and presents a meta-algorithm for computing rigid graph alignments. The method is versatile, in that it admits a range of topological and structural alignment techniques into an expectation maximization framework. The paper presents a detailed experimental study demonstrating significant performance improvements from rigid graph alignments over prior methods. Finally, in the context of human brain connectomes, it demonstrates significant improvements in distinguishability of connectomes and identification of individual signatures in brains. Our results open significant new avenues of research in the area of rigid graph aligners, akin to topological aligners two decades back.


\balance

\bibliographystyle{abbrv}
\bibliography{references}

\begin{thebibliography}{10}

\bibitem{Barabasi99}
A.-L. Barab{\'a}si and R.~Albert.
\newblock Emergence of scaling in random networks.
\newblock {\em Science}, 286(5439):509--512, 1999.

\bibitem{Bayati09}
M.~Bayati, D.~F. Gleich, A.~Saberi, and Y.~Wang.
\newblock Message-passing algorithms for sparse network alignment.
\newblock {\em ACM Trans. Knowl. Discov. Data}, 7(1):3:1--3:31, Mar. 2013.

\bibitem{Lassig04}
J.~Berg and M.~L{\"a}ssig.
\newblock Local graph alignment and motif search in biological networks.
\newblock {\em Proceedings of the National Academy of Sciences},
  101(41):14689--14694, 2004.

\bibitem{Berman00}
H.~M. Berman, J.~Westbrook, Z.~Feng, G.~Gilliland, T.~N. Bhat, H.~Weissig,
  I.~N. Shindyalov, and P.~E. Bourne.
\newblock The protein data bank.
\newblock {\em Nucleic acids research}, 28(1), 2000.

\bibitem{Besl92}
P.~Besl and H.~McKay.
\newblock A method for registration of 3-d shapes. ieee trans pattern anal mach
  intell.
\newblock {\em Pattern Analysis and Machine Intelligence, IEEE Transactions
  on}, 14:239--256, 03 1992.

\bibitem{Bouaziz13}
S.~Bouaziz, A.~Tagliasacchi, and M.~Pauly.
\newblock Sparse iterative closest point.
\newblock In {\em Proceedings of the Eleventh Eurographics/ACMSIGGRAPH
  Symposium on Geometry Processing}, SGP '13, pages 113--123, Aire-la-Ville,
  Switzerland, Switzerland, 2013. Eurographics Association.

\bibitem{Buneman16}
P.~Buneman and S.~Staworko.
\newblock Rdf graph alignment with bisimulation.
\newblock In {\em International Conference on Very Large Databases (VLDB)},
  volume~9 of {\em Proceedings of the VLDB Endowment}, pages 1149--1160, 2016.

\bibitem{Chetverikov02}
D.~Chetverikov, D.~Svirko, D.~Stepanov, and P.~Krsek.
\newblock The trimmed iterative closest point algorithm.
\newblock In {\em Pattern Recognition, 2002. Proceedings. 16th International
  Conference on}, volume~3, pages 545--548, USA, 2002. IEEE.

\bibitem{Ciriello12}
G.~Ciriello, M.~Mina, P.~H. Guzzi, M.~Cannataro, and C.~Guerra.
\newblock Alignnemo: A local network alignment method to integrate homology and
  topology.
\newblock {\em PLOS ONE}, 7(6):1--14, 06 2012.

\bibitem{Conroy11}
B.~R. Conroy and P.~J. Ramadge.
\newblock The grouped two-sided orthogonal procrustes problem.
\newblock In {\em 2011 IEEE International Conference on Acoustics, Speech and
  Signal Processing (ICASSP)}, pages 3688--3691, May 2011.

\bibitem{Eggert97}
D.~Eggert, A.~Lorusso, and R.~Fisher.
\newblock Estimating 3-d rigid body transformations: a comparison of four major
  algorithms.
\newblock {\em Machine Vision and Applications}, 9(5):272--290, Mar 1997.

\bibitem{Emmert16}
F.~Emmert-Streib, M.~Dehmer, and Y.~Shi.
\newblock Fifty years of graph matching, network alignment and network
  comparison.
\newblock {\em Inf. Sci.}, 346(C):180--197, June 2016.

\bibitem{VanEssen13}
D.~C.~V. Essen, S.~M. Smith, D.~M. Barch, T.~E. Behrens, E.~Yacoub, and
  K.~Ugurbil.
\newblock The wu-minn human connectome project: An overview.
\newblock {\em NeuroImage}, 80:62 -- 79, 2013.

\bibitem{VanEssen12}
D.~V. Essen, K.~Ugurbil, E.~Auerbach, D.~Barch, T.~Behrens, R.~Bucholz,
  A.~Chang, L.~Chen, M.~Corbetta, S.~Curtiss, S.~D. Penna, D.~Feinberg,
  M.~Glasser, N.~Harel, A.~Heath, L.~Larson-Prior, D.~Marcus, G.~Michalareas,
  S.~Moeller, R.~Oostenveld, S.~Petersen, F.~Prior, B.~Schlaggar, S.~Smith,
  A.~Snyder, J.~Xu, and E.~Yacoub.
\newblock The human connectome project: A data acquisition perspective.
\newblock {\em NeuroImage}, 62(4):2222 -- 2231, 2012.

\bibitem{Feizi16}
S.~Feizi, G.~Quon, M.~R. Mendoza, M.~M{\'{e}}dard, M.~Kellis, and A.~Jadbabaie.
\newblock Spectral alignment of networks.
\newblock {\em CoRR}, abs/1602.04181, 2016.

\bibitem{Finn15}
E.~S. Finn, X.~Shen, D.~Scheinost, M.~D. Rosenberg, J.~Huang, M.~M. Chun,
  X.~Papademetris, and R.~T. Constable.
\newblock Functional connectome fingerprinting: identifying individuals using
  patterns of brain connectivity.
\newblock {\em Nature Neuroscience}, 18(11), 2015.

\bibitem{Gilbert59}
E.~N. Gilbert.
\newblock Random graphs.
\newblock {\em Ann. Math. Statist.}, 30(4):1141--1144, 12 1959.

\bibitem{Heimann18}
M.~Heimann, W.~Lee, S.~Pan, K.-Y. Chen, and D.~Koutra.
\newblock Hashalign: Hash-based alignment of multiple graphs.
\newblock In D.~Phung, V.~S. Tseng, G.~I. Webb, B.~Ho, M.~Ganji, and
  L.~Rashidi, editors, {\em Advances in Knowledge Discovery and Data Mining},
  pages 726--739, Cham, 2018. Springer International Publishing.

\bibitem{Heimann18a}
M.~Heimann, H.~Shen, T.~Safavi, and D.~Koutra.
\newblock Regal.
\newblock {\em Proceedings of the 27th ACM International Conference on
  Information and Knowledge Management - CIKM ’18}, 2018.

\bibitem{Jenkinson02}
M.~Jenkinson, P.~Bannister, M.~Brady, and S.~Smith.
\newblock Improved optimization for the robust and accurate linear registration
  and motion correction of brain images.
\newblock {\em NeuroImage}, 17(2):825 -- 841, 2002.

\bibitem{Kabsch76}
W.~Kabsch.
\newblock {A solution for the best rotation to relate two sets of vectors}.
\newblock {\em Acta Crystallographica Section A}, 32(5):922--923, Sep 1976.

\bibitem{Khan12}
A.~M. Khan, D.~F. Gleich, A.~Pothen, and M.~Halappanavar.
\newblock A multithreaded algorithm for network alignment via approximate
  matching.
\newblock In {\em Proceedings of the International Conference on High
  Performance Computing, Networking, Storage and Analysis}, SC '12, pages
  64:1--64:11, Los Alamitos, CA, USA, 2012. IEEE Computer Society Press.

\bibitem{Klau09}
G.~W. Klau.
\newblock A new graph-based method for pairwise global network alignment.
\newblock {\em BMC Bioinformatics}, 10(1):S59, Jan 2009.

\bibitem{Kollias12}
G.~Kollias, S.~Mohammadi, and A.~Grama.
\newblock Network similarity decomposition (nsd): A fast and scalable approach
  to network alignment.
\newblock {\em IEEE Trans. on Knowl. and Data Eng.}, 24(12):2232--2243, Dec.
  2012.

\bibitem{Kuchaiev10}
O.~Kuchaiev, T.~Milenkovi{\'c}, V.~Memi{\v s}evi{\'c}, W.~Hayes, and N.~Pr{\v
  z}ulj.
\newblock Topological network alignment uncovers biological function and
  phylogeny.
\newblock {\em Journal of The Royal Society Interface}, 2010.

\bibitem{Liao09}
C.-S. Liao, K.~Lu, M.~Baym, R.~Singh, and B.~Berger.
\newblock Isorankn: spectral methods for global alignment of multiple protein
  networks.
\newblock {\em Bioinformatics}, 25(12):i253--i258, 2009.

\bibitem{Melnik02}
S.~Melnik, H.~Garcia-Molina, and E.~Rahm.
\newblock Similarity flooding: a versatile graph matching algorithm and its
  application to schema matching.
\newblock In {\em Proceedings 18th International Conference on Data
  Engineering}, pages 117--128. IEEE, 2002.

\bibitem{Mohammadi17}
S.~Mohammadi, D.~F. Gleich, T.~G. Kolda, and A.~Grama.
\newblock Triangular alignment tame: A tensor-based approach for higher-order
  network alignment.
\newblock {\em IEEE/ACM Trans. Comput. Biol. Bioinformatics}, 14(6):1446--1458,
  Nov. 2017.

\bibitem{Murphy09}
K.~Murphy, R.~M. Birn, D.~A. Handwerker, T.~B. Jones, and P.~A. Bandettini.
\newblock The impact of global signal regression on resting state correlations:
  Are anti-correlated networks introduced?
\newblock {\em NeuroImage}, 44(3):893--905, 2009.

\bibitem{Nassar18}
H.~Nassar, G.~Kollias, A.~Grama, and D.~F. Gleich.
\newblock Low rank methods for multiple network alignment, 2018.

\bibitem{Patro12}
R.~Patro and C.~Kingsford.
\newblock Global network alignment using multiscale spectral signatures.
\newblock {\em Bioinformatics}, 28(23):3105--3114, 2012.

\bibitem{Riederer16}
C.~Riederer, Y.~Kim, A.~Chaintreau, N.~Korula, and S.~Lattanzi.
\newblock Linking users across domains with location data: Theory and
  validation.
\newblock In {\em Proceedings of the 25th International Conference on World
  Wide Web}, WWW '16, pages 707--719, Republic and Canton of Geneva,
  Switzerland, 2016. International World Wide Web Conferences Steering
  Committee.

\bibitem{Rusinkiewicz01}
S.~Rusinkiewicz and M.~Levoy.
\newblock Efficient variants of the icp algorithm.
\newblock {\em Proc. 3DIM}, 2001, 10 2001.

\bibitem{Sabata91}
B.~Sabata and J.~Aggarwal.
\newblock Estimation of motion from a pair of range images: A review.
\newblock {\em CVGIP: Image Understanding}, 54(3):309 -- 324, 1991.

\bibitem{Schoenemann66}
P.~H. Sch{\"o}nemann.
\newblock A generalized solution of the orthogonal procrustes problem.
\newblock {\em Psychometrika}, 31(1):1--10, Mar 1966.

\bibitem{Singh07b}
R.~Singh, J.~Xu, and B.~Berger.
\newblock Pairwise global alignment of protein interaction networks by matching
  neighborhood topology.
\newblock In T.~Speed and H.~Huang, editors, {\em Research in Computational
  Molecular Biology}, pages 16--31, Berlin, Heidelberg, 2007. Springer Berlin
  Heidelberg.

\bibitem{Singh07a}
R.~Singh, J.~Xu, and B.~Berger.
\newblock Pairwise global alignment of protein interaction networks by matching
  neighborhood topology.
\newblock In {\em Proceedings of the 11th Annual International Conference on
  Research in Computational Molecular Biology}, RECOMB'07, pages 16--31,
  Berlin, Heidelberg, 2007. Springer-Verlag.

\bibitem{Smith02}
S.~M. Smith.
\newblock Fast robust automated brain extraction.
\newblock {\em Human Brain Mapping}, 17(3):143 -- 155, November 2002.

\bibitem{Smith13}
S.~M. Smith, C.~F. Beckmann, J.~Andersson, E.~J. Auerbach, J.~Bijsterbosch,
  G.~Douaud, E.~Duff, D.~A. Feinberg, L.~Griffanti, M.~P. Harms, M.~Kelly,
  T.~Laumann, K.~L. Miller, S.~Moeller, S.~Petersen, J.~Power,
  G.~Salimi-Khorshidi, A.~Z. Snyder, A.~T. Vu, M.~W. Woolrich, J.~Xu,
  E.~Yacoub, K.~Uğurbil, D.~C.~V. Essen, and M.~F. Glasser.
\newblock Resting-state {fMRI} in the human connectome project.
\newblock {\em NeuroImage}, 80:144 -- 168, 2013.

\bibitem{Sussenguth}
E.~H. Sussenguth.
\newblock A graph-theoretic algorithm for matching chemical structures.
\newblock {\em Journal of Chemical Documentation}, 5(1):36--43, 1965.

\bibitem{Umeyama88}
S.~Umeyama.
\newblock An eigendecomposition approach to weighted graph matching problems.
\newblock {\em IEEE Transactions on Pattern Analysis and Machine Intelligence},
  10(5):695--703, Sep. 1988.

\bibitem{Vijayan18}
V.~Vijayan and T.~Milenkovic.
\newblock Multiple network alignment via multimagna++.
\newblock {\em IEEE/ACM Trans. Comput. Biol. Bioinformatics}, 15(5):1669--1682,
  Sept. 2018.

\bibitem{Zelinka75}
B.~Zelinka.
\newblock On a certain distance between isomorphism classes of graphs.
\newblock {\em Časopis pro p̌est.Mathematiky}, 100(4):371--373, 1975.

\bibitem{Zelinka92}
B.~Zelinka.
\newblock Distances between graphs (extended abstract).
\newblock In {\em Fourth Czechoslovakian Symposium on Combinatorics, Graphs and
  Complexity}, volume~51 of {\em Annals of Discrete Mathematics}, pages 355 --
  361. Elsevier, 1992.

\bibitem{Zhang16}
S.~Zhang and H.~Tong.
\newblock Final: Fast attributed network alignment.
\newblock In {\em Proceedings of the 22nd ACM SIGKDD International Conference
  on knowledge discovery and data mining}, KDD '16, pages 1345--1354. ACM,
  2016.

\end{thebibliography}
\bibliographystyle{plainnat}

\end{document}